\documentclass[journal=jacsat,manuscript=article,layout=twocolumn]{achemso}

\usepackage[version=4]{mhchem}
\usepackage[utf8]{inputenc}
\usepackage{graphicx}
\usepackage{epsfig}
\usepackage{epstopdf}
\usepackage{float}
\usepackage{amsmath}
\usepackage{caption}
\usepackage{subcaption}
\usepackage{natbib}
\usepackage{xcolor}
\usepackage{csquotes}
\usepackage{caption}
\title{Chiral Phonons in Graphyne \\}

\author{Subhendu Mishra}
\affiliation{Materials Research Centre, Indian Institute of Science, Bangalore 560012, India}
\author{Arpan Chakraborty}
\affiliation{Materials Research Centre, Indian Institute of Science, Bangalore 560012, India}
\author{Douglas S Galvao}
\email{galvao@ifi.unicamp.br}
\affiliation{Department of Applied Physics and Center for Computational Engineering and Sciences, State
University of Campinas, Campinas, 13083-859, SP, Brazil}
\author{Pedro A. S. Autreto}
\affiliation{Center for Natural and Human Sciences (CCNH), Federal University of ABC Rua Santa Adélia 166, Santo André 09210-170, Brazil.}
\email{pedro.autreto@uafbc.edu.br}
\author{Abhishek Kumar Singh}
\email{abhishek@iisc.ac.in}
\affiliation{Materials Research Centre, Indian Institute of Science, Bangalore 560012, India}

\begin{document}

\maketitle

\begin{abstract}
Chiral phonons, quantized lattice vibrations with circular polarization and non-zero angular momentum, offer new perspectives for phononic and quantum device engineering. Graphyne could be a promising candidate due to its unique lattice geometry, valley-structured electronic bands, and thermal transport capabilities. However, chiral phonons in graphyne remain unexplored owing to the existence of inversion ($\mathcal{P}$) and time-reversal ($\mathcal{T}$) symmetries. Herein, we have demonstrated the existence of chiral phonons in graphynes, achieved by breaking combined $\mathcal{PT}$ symmetry through atomic-selective substitutional doping. We find that the B, N, dopants and ortho BN co-dopant in 6-6-12 and $\gamma$-graphynes induce localized structural deformations. These deformations lift phonon degeneracies away from $\Gamma$ point and give rise to circularly polarized vibrational modes. We further established a strong correlation between chiral phonon angular momentum and electron affinity of dopants. Electron-rich dopants increase local electron density which could enable chiral phonon modes to couple more effectively with electronic environment. This in turn increases phonon angular momentum, indicating potential role of electron-phonon interactions in angular momentum modulation of chiral phonons. Our prosposed approach provides a tunable route for controlling chiral phonon behavior, paving way for development of advanced phononic devices.

\end{abstract} 

\section{Keywords}
\textit{Graphyne, Chiral Phonon, Phonon Polarization, Phonon Angular Momentum Modulation, Electron-Phonon Interaction}

\maketitle
\section{Introduction}

Phonons, the quantized lattice vibrations in solids, are central for understanding thermal, mechanical, and electronic behavior of materials. Generally, phonon modes are considered to be linearly polarized and carry no angular momentum. However, theoretical predictions~\cite{zhang2015chiral} and experimental observations~\cite{zhu2018observation,ueda2023chiral,yin2021chiral} have uncovered the existence of vibrational modes with circular polarization and non-zero angular momentum. The circularly polarized modes, known as chiral phonons, have recently gained attention for their ability to control valley, spin, and heat transport in 2D materials. For example, in transition metal dichalcogenides, such as WSe$_2$ and MoS$_2$, broken inversion symmetry ($\mathcal{P}$) and strong spin-orbit coupling give rise to chiral phonons at K (K$^{'}$) valleys in the Brillouin zone (BZ) ~\cite{zhu2018observation,maity2022chiral}. These chiral phonons selectively couple to valley electrons and other quasiparticles, influencing intervalley scattering, phonon magnetization, and various electronic and optical phenomena~\cite{chen2015helicity,kang2018holstein,li2019emerging,li2019momentum,luo2023large,chaudhary2024giant,carvalho2017intervalley}. The coupling between chiral phonons and intervalley excitons could enhance exciton lifetimes while preserving valley polarization~\cite{mishra2024symmetry,barik2024valley}, an essential feature for valleytronic applications~\cite{li2019emerging}. Moreover, coupling of phonon angular momentum with electron spin enables spin current generation. This phenomenon is known as spin Seebeck effect, which highlights the potential of chiral phonons for spin caloritronic applications~\cite{ohe2024chirality,uchida2008observation,kim2023chiral,pols2025chiral,zhang2012molecular}.

Graphynes \cite{Ray} (Figure \ref{fig:fig-1}) are a structurally diverse family of carbon allotropes with sp and sp$^2$ hybridized atoms.
They exhibit graphene-like electronic and mechanical properties, along with advantages like non-zero band gaps.~\cite{shao2015optical,peng2012mechanical,majidi2023comparative}. $\gamma$-graphyne has been recently synthesized in large quantities \cite{Valentin1, Valentin2}. Some graphyne structures exhibit anisotropic lattice geometries, valley-structured electronic bands, and tunable Dirac-like dispersion~\cite{malko2012competition,majidi2023comparative}, which are favorable for hosting chiral phonons. Additionally, graphyne offers moderate thermal transport and mechanical flexibility, making it suitable for phononic and quantum device applications.~\cite{zhang2012molecular,ali2025exploration}. Despite a promising theoretical landscape, realization of chiral phonons in graphyne remains unexplored, primarily due to existence of inversion symmetry in many of its pristine allotropes. 

Here, we propose atom-selective substitutional doping as a symmetry-breaking mechanism that can induce phonon chirality in graphyne. Undoped 6-6-12- and $\gamma$-graphyne (Figure \ref{fig:fig-1}) do not exhibit phonon chirality due to their inherent inversion symmetry. However, doping with B, N, and co-doping with BN introduce localized structural deformations, leading to broken $\mathcal{PT}$ symmetry. This symmetry breaking lifts phonon degeneracies, enabling circularly polarized chiral phonon modes. A strong correlation is observed between chiral phonon angular momentum and the electron affinity of the dopants. Electron-rich dopants increase local electron density, which could allow phonon modes to interact more effectively with electronic environment. This leads to enhanced phonon angular momentum, indirectly highlights role of electron-phonon interaction in modulating angular momentum of vibrational modes. Our study presents a controllable approach for activating chiral phonons and positions graphyne as a strong candidate for phonon-based quantum devices.


\section{Results and Discussions}

\begin{figure*}[htb!]
\begin{center}
\includegraphics[width=0.9\linewidth]{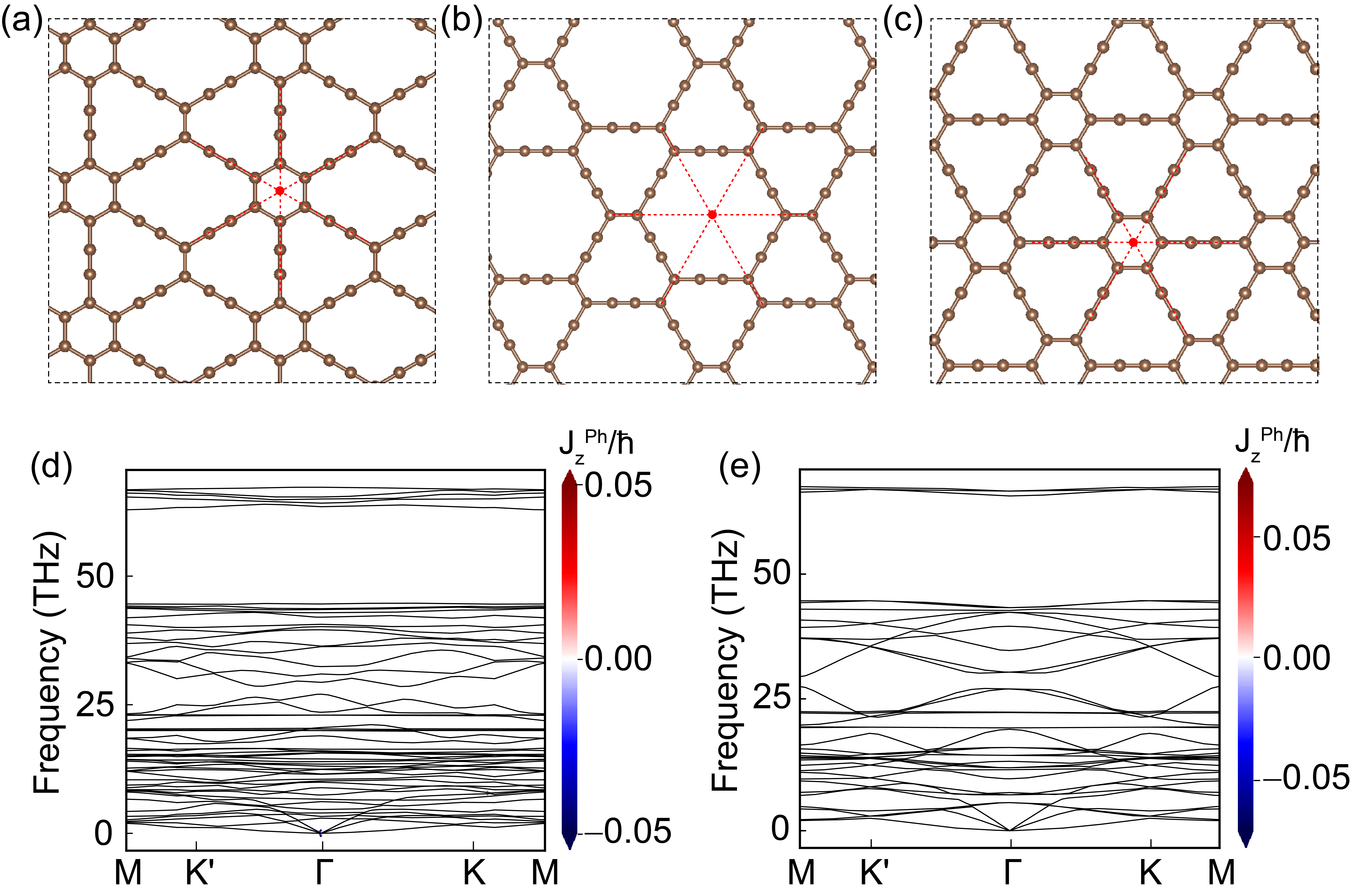}
\end{center}
\caption{Top-views of the optimized structures (a-c) 6-6-12, $\beta$,  and $\gamma$-graphyne, respectively. The red color solid circle indicates the inversion center of the structures. (d-e) Phonon band structures of 6-6-12- and $\gamma-$graphyne, respectively. The color bar indicates the values of $J{_z^{ph}}$. The positive and negative values, represented by red and blue colors, respectively, indicate the right- and left-circularly polarized phonons. The negligible value of $J{_z^{ph}}$ indicates that the phonon modes do not exhibit any chirality for both 6-6-12,and $\gamma$-graphyne.}
\label{fig:fig-1}
\end{figure*}

Figure \ref{fig:fig-1}a shows the optimized 6-6-12-graphyne monolayer. It exhibits an orthogonal lattice with in-plane lattice parameters of $a=$9.43~\AA\ and $b=$6.91~\AA\ and internal angles of 90$^\circ$, forming a rectangular unit cell. Its framework is composed of repeating segments of two benzene rings ($6$-membered). They are connected by a $12$-membered carbon ring, resulting in a porous and directionally extended architecture. The framework integrates three distinct types of C-C bonds: linear C(sp)-C(sp) triple bonds ($\sim$1.22 \AA), bridging C(sp$^2$)-C(sp) bonds ($\sim$1.41 \AA), and aromatic C(sp$^2$)-C(sp$^2$) bonds ($\sim$1.43 \AA). Figure \ref{fig:fig-1}b shows optimized $\beta-$graphyne monolayer, having in-plane lattice constants of $\sim$9.48~\AA\ and internal angle of 120$^\circ$, exhibiting a trigonal symmetry. Its structure integrates both sp and sp$^2$ hybridized carbon atoms, forming a porous framework composed of benzene rings and acetylenic ($\ce{-C#C-}$) linkages. It features three distinct carbon-carbon bond types: aromatic C(sp$^2$)-C(sp$^2$) bonds measuring 1.46 \AA, bridging C(sp$^2$)-C(sp) bonds of 1.39 \AA, and short C(sp)-C(sp) triple bonds of 1.23 \AA. On the other hand, monolayer $\gamma$-graphyne, as shown in Figure \ref{fig:fig-1}c, is a 2D carbon allotrope with hexagonal symmetry (space group P6m).~\cite{Ray} It has an optimized in-plane lattice constant of 6.89 \AA, and forms a planar lattice with an internal angle of 120$^\circ$. It possesses a distinctive arrangement of sp- and sp$^2$-hybridized carbon atoms. This includes C(sp)-C(sp) triple bonds (1.22 \AA), C(sp$^2$)-C(sp$^2$) single linkages between aromatic rings and acetylene chains (1.41 \AA), and C(sp$^2$)-C(sp$^2$) bonds of benzene units (1.43 \AA). The hybrid bonding creates an open $\pi$-conjugated structure with tunable electronic and directional functionality distinct from that of graphene.

\begin{figure*}[htb!]
\begin{center}
\includegraphics[width=1\linewidth]{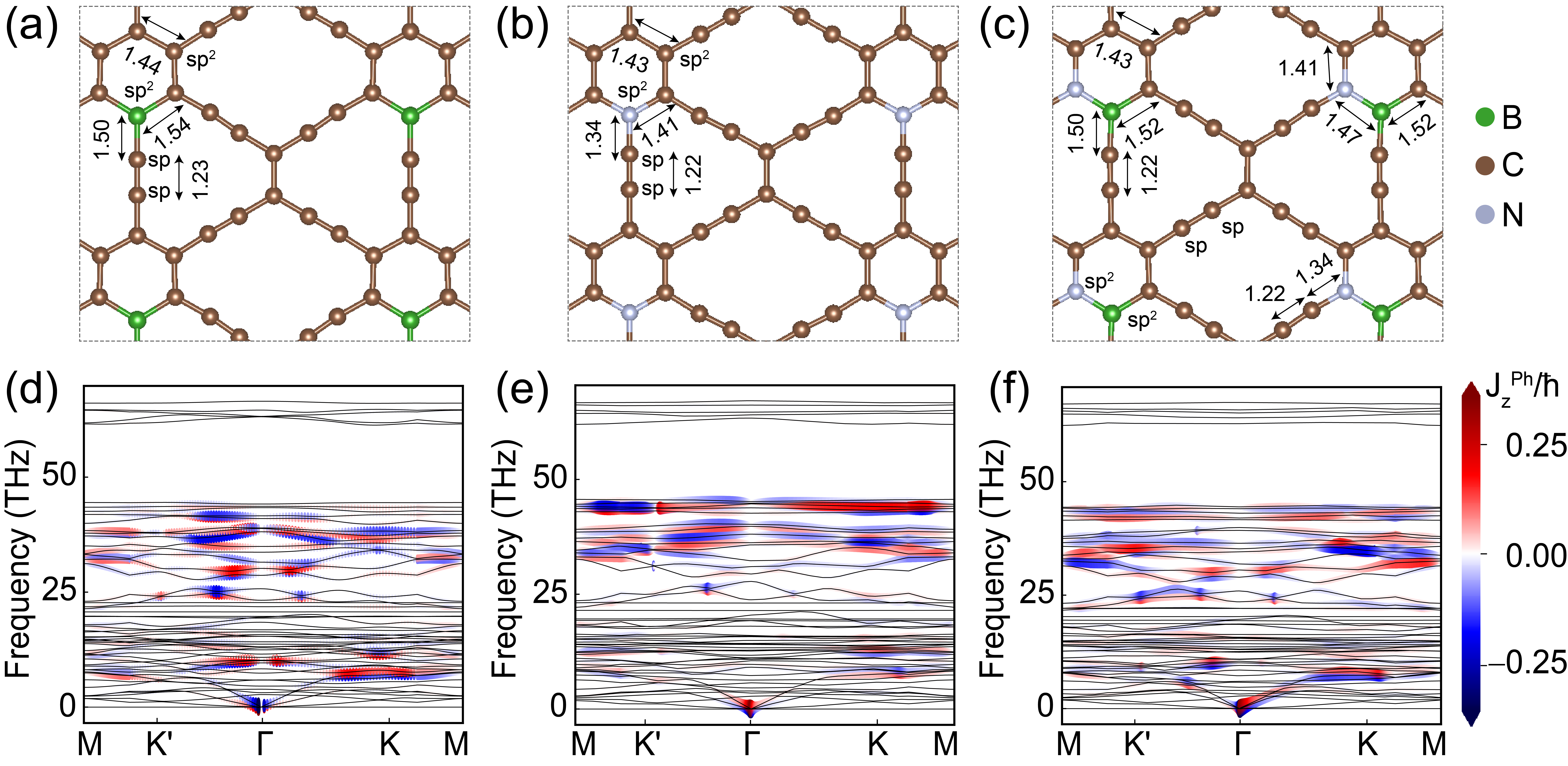}
\end{center}
\caption{Top-views of optimized structures of (a-c) B doped, N doped, and ortho BN co-doped 6-6-12-graphynes, respectively. The inversion symmetry in all of these systems is absent. (d-f) Corresponding phonon band structures with superimposed phonon angular momentum. The color bar indicates the values of $J{_z^{ph}}$. The positive and negative values, represented by red and blue colors, respectively, indicate the right- and left-circularly polarized phonons. The finite values of $J{_z^{ph}}$ indicate that the phonon modes exhibit chirality for all three doped graphyne structures.}
\label{fig:fig-2}
\end{figure*}

The phonon band structures of 6-6-12 and $\gamma-$graphynes are presented in Figure \ref{fig:fig-1}d,e, respectively. The absence of imaginary frequency modes in 6-6-12 and $\gamma-$graphynes suggests that they are dynamically stable. On the other hand, the presence of negative phonon frequency modes indicates that $\beta-$graphyne is dynamically unstable~\cite{perkgoz2014vibrational}, as shown in Figure S1 in supplementary information (SI). We therefore focus exclusively on phonon band structures of 6-6-12 and $\gamma$-graphynes. The phonon polarization vector characterizes the chirality of these phonon modes. The phonon polarization vector for the unit cell having N atoms is defined as~\cite{zhang2015chiral,zhang2014angular}:

\begin{equation}
\label{eq1}
\epsilon = \sum_{\alpha=1}^{N} \epsilon_{R_{\alpha}}|R_{\alpha}\rangle+\epsilon_{L_{\alpha}}|L_{\alpha}\rangle   
\end{equation}
The $|R_{\alpha}\rangle$ and $|L_{\alpha}\rangle$ are the basis vectors for the right-circular and left-circularly polarized phonons, respectively.
The phonon circular polarization operator along the $z$ axis is defined as:

\begin{equation}
\label{eq2}
S_z = \sum_{\alpha=1}^{N} |R_{\alpha}\rangle \langle R_{\alpha}|-|L_{\alpha}\rangle \langle L_{\alpha}|
=\begin{pmatrix}
0 & -i\\
i & 0\\ \end{pmatrix} \otimes I_{2\times2}
\end{equation}

The  corresponding phonon circular polarization is defined as:

\begin{equation}
\label{eq3}
 {s_z}^{ph}=(\epsilon^{+}S_z\epsilon)\hbar  
 = \sum_{\alpha=1}^{N} (|\epsilon_{R_{\alpha}}|^2-|\epsilon_{L_{\alpha}}|^2)\hbar
\end{equation}

The value of ${s_z}^{ph} >$ 0 indicates that the phonon is right-handed, whereas ${s_z}^{ph} <$ 0 indicates that the phonon is left-handed.  For ${s_z}^{ph} =$ 0, phonon modes are linear polarized, and chirality is absent. The ${s_z}^{ph}$ has the same form as that of phonon angular momentum~\cite{zhang2014angular}, defined as

\begin{equation}
\label{eq4}
{J_z}^{ph} = \left({\epsilon_{k}}^+ M \epsilon_{k} \right)\hbar = {s_z}^{ph}
\end{equation}
 where $M = \begin{bmatrix}
 0 & -i\\
 i & 0\\   
\end{bmatrix}\otimes I_{2\times2}$ is the mass matrix. 
We have calculated the phonon angular momentum using Eq.~\ref{eq4} and superimposed it onto the phonon dispersion, as shown in Figure \ref{fig:fig-1}d-e. A negligible phonon angular momentum is observed both for 6-6-12 and $\gamma$-graphynes. This happens because all of them exhibit inversion symmetry (\(\mathcal{P}\)) and also time-reversal symmetry (\(\mathcal{T}\)) in their pristine form. These symmetries impose strict constraints on the phonon angular momentum \(J^{\text{ph}}(q)\). Under $\mathcal{T}$, the phonon angular momentum transforms as \(J^{\text{ph}}(q) \rightarrow -J^{\text{ph}}(-q)\), making it an odd function in momentum space. While, $\mathcal{P}$ transforms \(J^{\text{ph}}(q) = J^{\text{ph}}(-q)\), making it an even function. The simultaneous existence of both the symmetries leads to a contradiction unless \(J^{\text{ph}}(q) = 0\) for all \(q\).

On the other hand, upon boron (B) doping on 6-6-12-graphyne, the local bonding environment becomes asymmetric due to its electron deficiency, leading to elongated B-C(sp$^2$) bonds (1.50-1.54 \AA). The neighboring C(sp$^2$)-C(sp$^2$) bonds also stretch to 1.44 \AA, as shown in Figure \ref{fig:fig-2}a. In contrast, nitrogen (N) doping introduces localized bond contraction, with N-C(sp$^2$) bonds shortening to 1.34 ~\AA\ and neighboring C(sp$^2$)-C(sp$^2$) bonds are slightly reduced to 1.43 \AA. The C(sp)-C(sp) triple bonds remain largely unaffected in both cases (1.22-1.23 \AA), preserving the rigidity of the acetylenic backbone (Figure \ref{fig:fig-2}b). This asymmetry in local bonding environments breaks the spatial symmetry $\mathcal{P}$ of the pristine 6-6-12-graphyne. This would give rise to non-zero phonon polarization, as evident from phonon band structures of B and N doped 6-6-12-graphynes, shown in Figure \ref{fig:fig-2}d,e, respectively. Red and blue colors represent positive and negative values of $J{_z^{ph}}$, corresponding to right- and left-circularly polarized phonons, respectively.

\begin{figure*}[htb!]
\begin{center}
\includegraphics[width=1\linewidth]{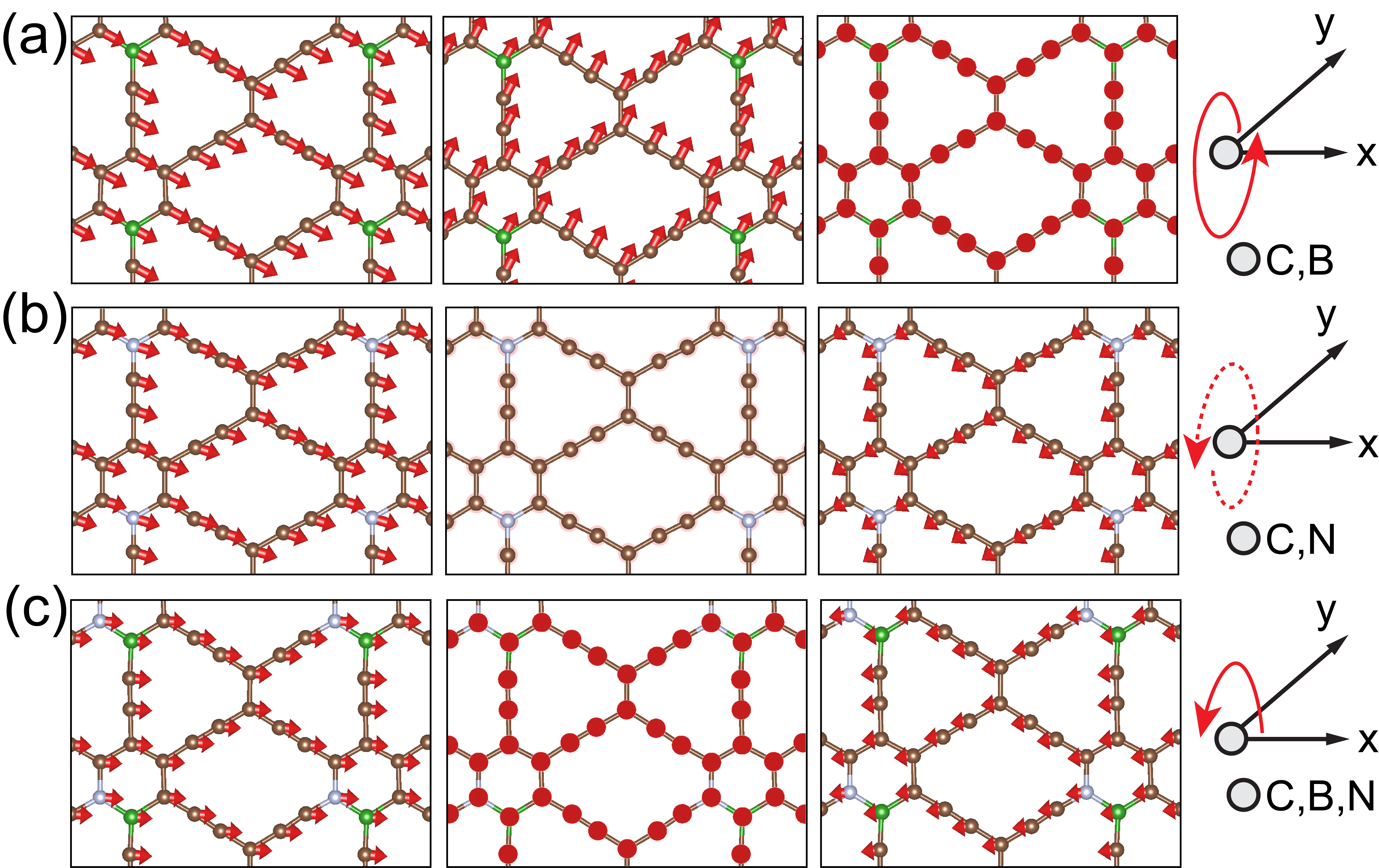}
\end{center}
\caption{Phonon eigenvector analysis at the $\Gamma$-point reveals that the superposition of degenerate vibrational modes induces circular atomic motion, resulting in finite phonon angular momentum ($J_z^{ph}$) in doped 6-6-12-graphyne structures. (a) For B doped 6-6-12-graphyne, the combination of A$^{'}$, A$^{''}$ and A$^{'}$, (b) for N doped, the superposition of three A$^{'}$ modes and (c) for ortho BN co-doped modes the superposition of A$^{''}$ and two A$^{'}$ results in non-zero $J_z^{ph}$ along the z-axis. Red arrows indicate the direction of phonon eigenvectors.}
\label{fig:fig-3}
\end{figure*}

The most thermodynamically stable ortho BN co-doped structure among all possible co-doped configurations (Figure S2,S3 in SI) of 6-6-12-graphyne is shown in Figure \ref{fig:fig-2}c. B and N atoms substitute adjacent carbon sites within a hexagonal ring, forming a compact dopant pair that promotes local symmetry and electronic balance. The optimized geometry reveals a slightly elongated B-C bond ($\sim$1.52 \AA), compared to N-C bond (1.34-1.41 \AA), consistent with larger atomic radius and lower electronegativity of B. The asymmetric bonding induces localized charge redistribution, with B as an electron acceptor and N as a donor, forming a built-in dipole and strain gradient. This breaks $\mathcal{P}$, giving rise to chiral phonon modes, as illustrated in Figure \ref{fig:fig-2}f.

While all these systems exhibit broken $\mathcal{P}$, they preserve $\mathcal{T}$, which would result combination of $\mathcal{P}$ and $\mathcal{T}$ i.e., $\mathcal{PT}$ symmetry to be broken. In such a scenario, the $J_z^{ph}$ will be non-zero~\cite{zhang2025measurement}. Additionally, all of them exhibit the rotoinversion symmetry (\(R_{\mathcal{P}} = C_{2z} + \mathcal{P}\)).
Under a spatial rotation \(R\), the angular momentum is transformed as:

\begin{equation}
J^{\text{ph}}(Rq) = \det(R)\, R J^{\text{ph}}(q),
\end{equation}

In the presence of $R_{\mathcal{P}}$, the wavevector transforms as \(q \rightarrow -Rq\) with $R=C_{2z}$, leading to \(J^{\text{ph}}(-Rq) = -R J^{\text{ph}}(q)\) as for two-fold rotations \(\det(R) = -1\). The coexistence of $R_{\mathcal{P}}$ and broken $\mathcal{PT}$ would result in non-zero phonon angular momentum, consistent with previous studies~\cite{zhang2025measurement,yang2025catalogue,zhang2025chirality}.

Moreover, due to the breaking of $\mathcal{P}$, the degeneracy of phonon branches is lifted at the high symmetry k points, away from the $\Gamma$ point. Our $\Gamma$-point phonon eigenvector analysis further shows that the superposition of degenerate modes in these doped systems could result in chiral phonon modes. For B doped 6-6-12-graphyne, the superposition of two $A^{'}$ and one $A^{''}$ vibrational modes, result in non-zero $J{_z^{ph}}$, as shown in Figure \ref{fig:fig-3}a. The red solid arrows on each atom represent the direction of phonon eigenvectors. The in-plane mode of vibration in the $x-y$ plane results in non-zero phonon angular momentum along the $z$ direction. For N doped 6-6-12-graphyne, superposition of three $A^{'}$ modes (Figure \ref{fig:fig-3}b) and for BN co-doped ortho 6-6-12-graphyne, superposition of $A^{''}$ and two $A^{'}$ modes (Figure \ref{fig:fig-3}c) gives rise to chiral phonon modes.

\begin{figure*}[htb!]
\begin{center}
\includegraphics[width=1\linewidth]{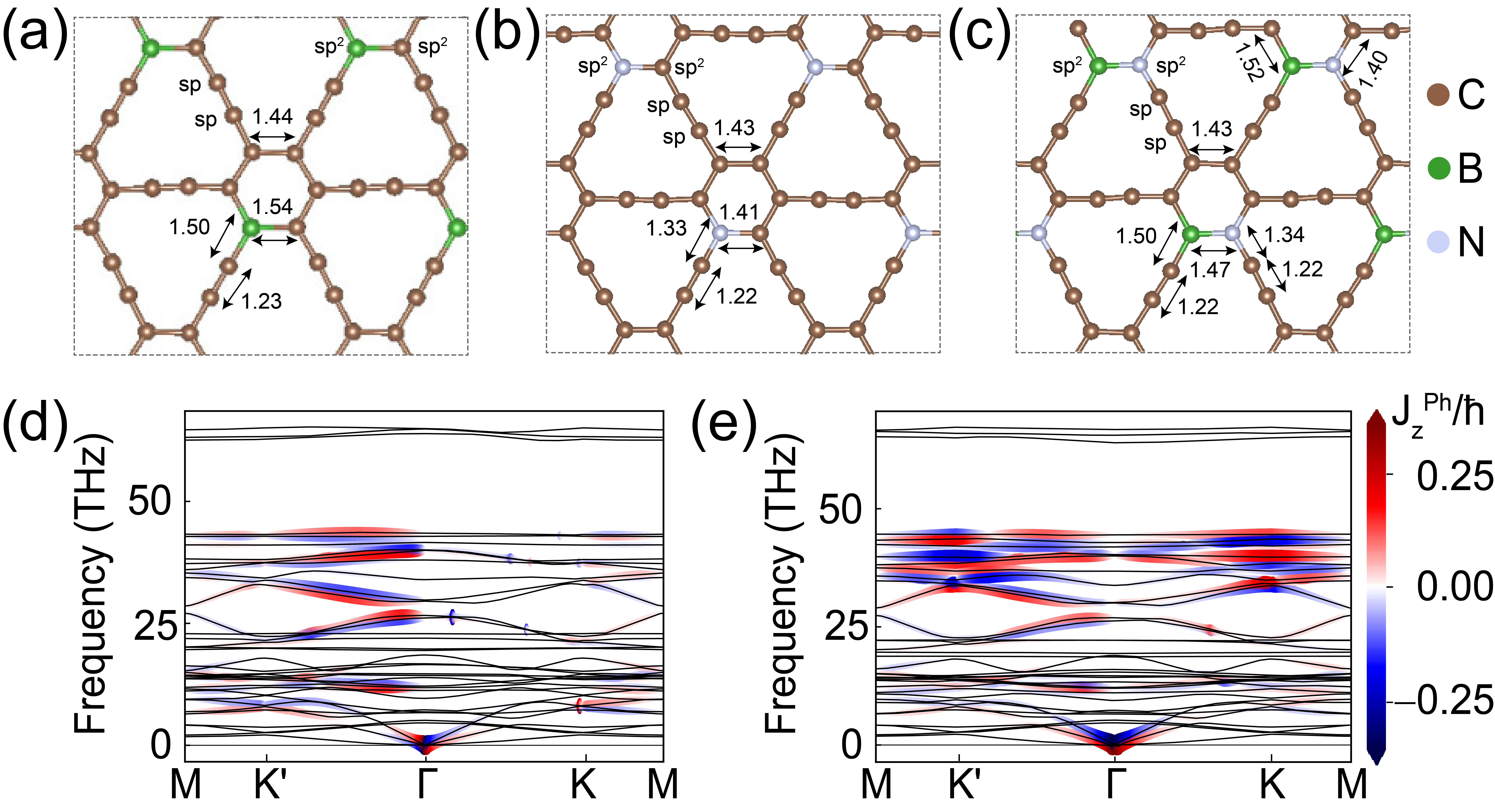}
\end{center}
\caption{Top-views of the optimized crystal structures of (a-c) B doped, N doped, and ortho BN co-doped $\gamma$-graphynes, respectively. Phonon band structures of (d-e) B doped and ortho BN co-doped $\gamma$-graphynes, respectively. The color bar indicates the values of phonon angular momentum ($J{_z^{ph}}$). The positive and negative values, represented by red and blue colors, respectively, indicate the right- and left-circularly polarized phonons. The finite values of $J{_z^{ph}}$ indicate that the phonon modes exhibit chirality for all these graphyne structures.}
\label{fig:fig-4}
\end{figure*}

On the other hand, although undoped $\gamma$-graphyne (Figure \ref{fig:fig-1}c) exhibits ideal hexagonal symmetry, B doping disrupts this balance due to its electron deficiency. This leads to anisotropic lattice expansion from 6.97 to 7.05 \AA. Moreover, B-C(sp$^2$) bonds stretch to 1.54 \AA, C(sp$^2$)-C(sp) bonds to 1.50 \AA, and aromatic C(sp$^2$)-C(sp$^2$) bonds to 1.44 \AA~(Figure \ref{fig:fig-4}a). These non-uniform bond elongations would break the local symmetry and result in chiral phonon modes, as shown in Figure \ref{fig:fig-4}d.

\begin{figure*}[htb!]
\begin{center}
\includegraphics[width=1\linewidth]{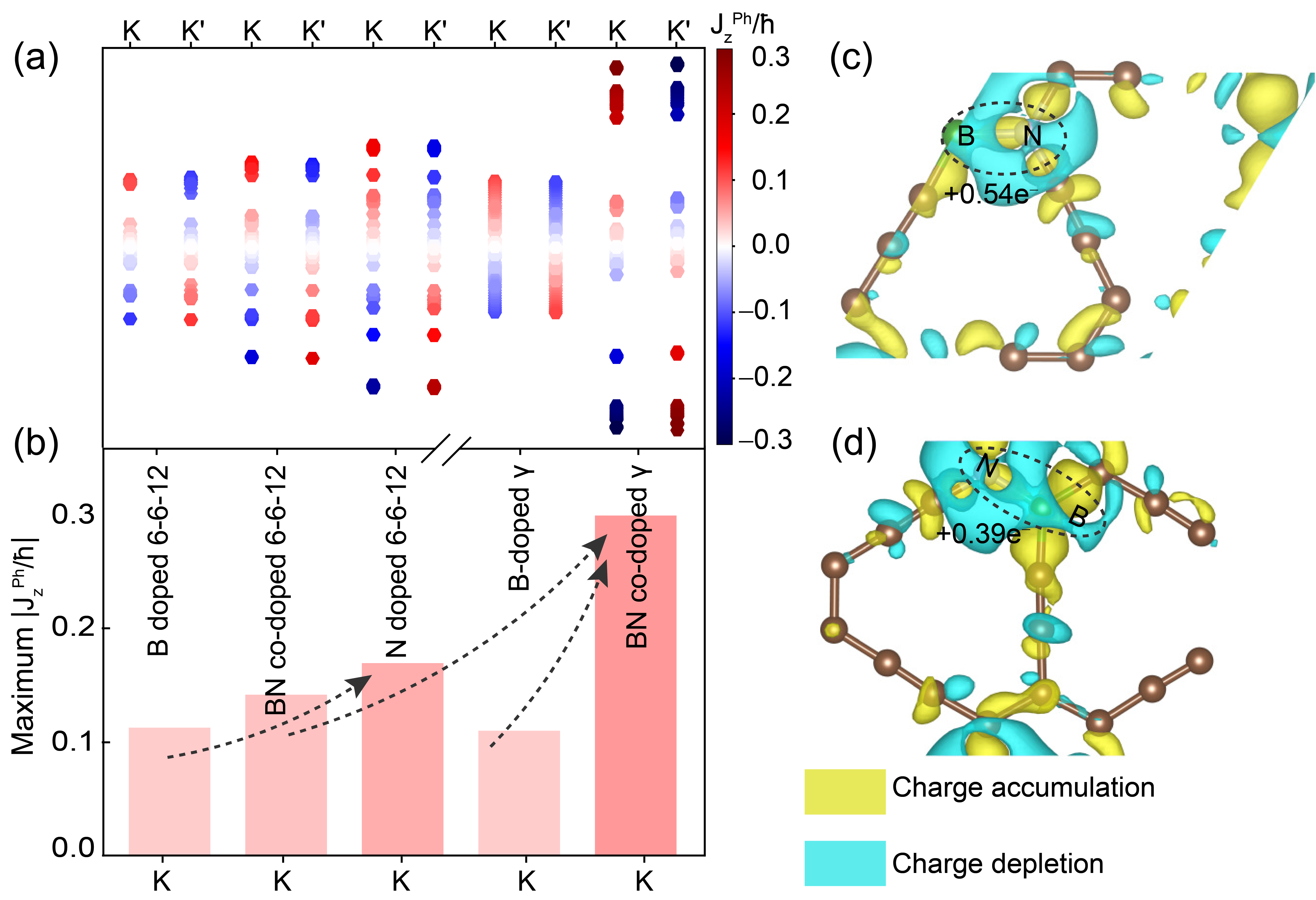}
\end{center}
\caption{(a) Phonon angular momentum ($J{_z^{ph}}$) values at the K and K$^{'}$ points for several doping configurations in 6-6-12 and $\gamma$-graphyne, illustrating the modulation of phonon chirality through doping. (b) The variation of maximum $|J{_z^{ph}}|$ values at K(K$^{'}$) points across different doping configurations. (c) The charge density difference plot in BN co-doped $\gamma$-graphyne, attributed to a large charge transfer of $+$0.54 e$^{-}$ from BN to the local environment. (d) The charge density difference plot of BN co-doped 6-6-12-graphyne, attributed to a charge transfer of $+$0.39 e$^{-}$ from BN to the local environment, resulting in relatively lower phonon angular momentum values compared to $\gamma$-graphyne. The yellow and cyan colors indicate charge accumulation and depletion, respectively. These results underscore the role of dopant-induced charge redistribution in tuning phonon angular momentum and chirality.}

\label{fig:fig-5}
\end{figure*}
\begin{figure*}[htb!]
\begin{center}
\includegraphics[width=0.95\linewidth]{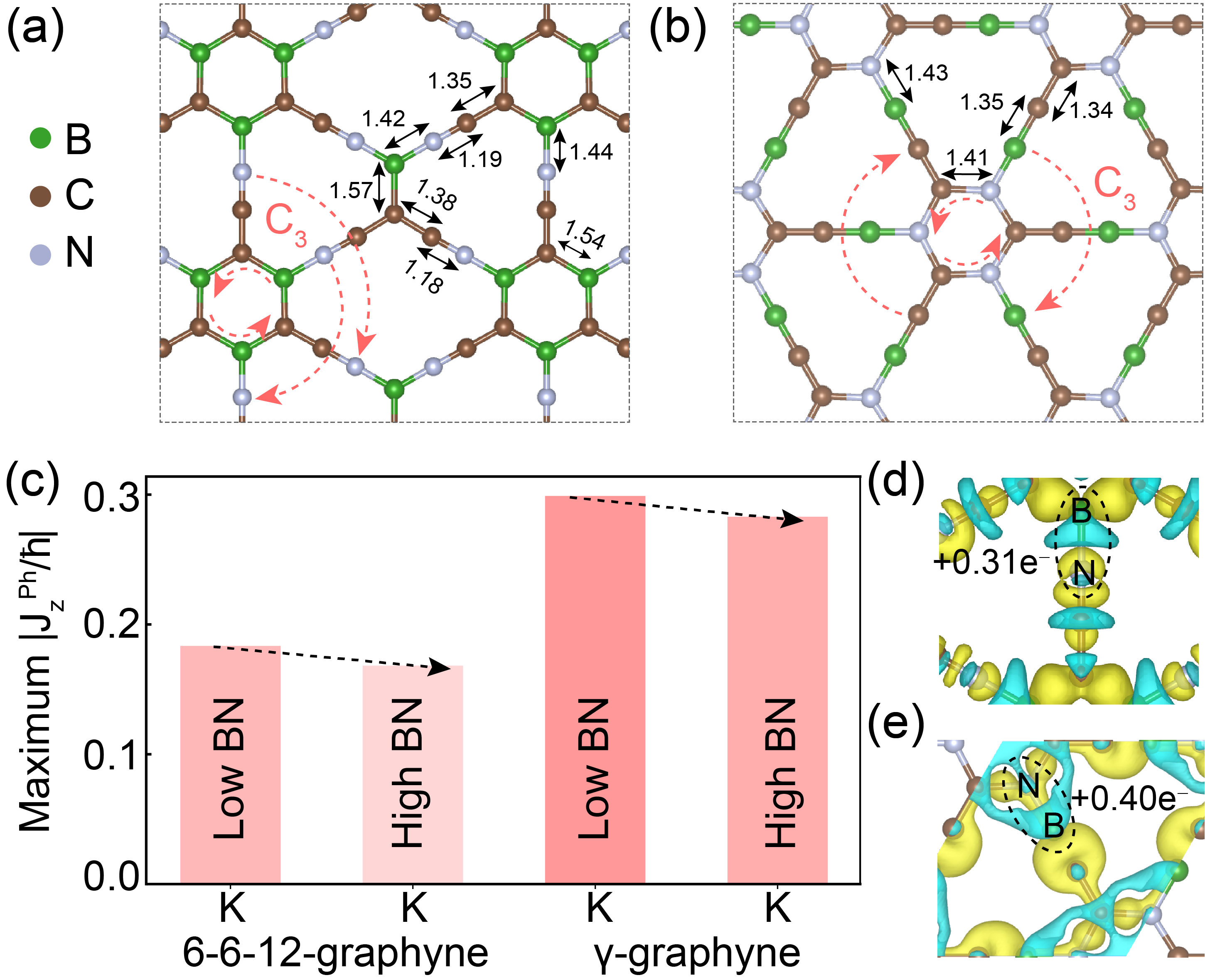}
\end{center}
\caption{(a) The optimized crystal structures of (a) 6-6-12 and (b) $\gamma$-graphynes with higher BN co-dopants. The bond length values are indicated in the inset of the figures. The red dotted arrow represents the presence of three-fold rotational symmetry ($C_3$). (c) The comparison of maximum phonon angular momentum $|J{_z^{ph}}|$ values at the K point for higher and lower BN co-doped 6-6-12 and $\gamma$-graphynes. The charge density difference plots of (d) 6-6-12 and (e) $\gamma$-graphynes with higher BN dopants. The yellow and cyan colors indicate charge accumulation and depletion, respectively. A charge transfer of 0.40 $e^{-}$ is observed from BN to the local environment in $\gamma$-graphyne, resulting in relatively higher $|J{_z^{ph}}|$ values compared to 6-6-12-grahyne, where charge transfer is 0.31 $e^{-}$. }
\label{fig:fig-6}
\end{figure*}

In contrast, N doping induces lattice contraction from 6.84 to 6.80 \AA, driven by contracted local bonding. This is reflected in shortened N-C(sp$^2$) bonds (1.41 \AA) and C(sp$^2$)-C(sp) bonds (1.33 \AA), while the triple bond length remains nearly unchanged ($\sim$1.22 \AA) (Figure \ref{fig:fig-4}b). The resulting contraction could introduce a non-zero value of $J{_z^{ph}}$. However, phonon band structure calculations reveal negative frequencies in N doped $\gamma$-graphyne (Figure S4 in SI), indicating the system is dynamically unstable. Consequently, this configuration was excluded from further analysis.

Ortho BN co-doping introduces even greater anisotropy, with structural changes highly sensitive to dopant configuration. In the ortho configuration, the lattice expands to 7.01 \AA. The B-C bond lengths range from 1.50 to 1.52 \AA, while N-C bonds contract to 1.35 \AA~(Figure \ref{fig:fig-4}c), resulting in pronounced bond polarization and symmetry breaking. These structural asymmetries create favorable conditions for the emergence of chiral phonon modes, as shown in Figure \ref{fig:fig-4}e. Similar to doped 6-6-12-graphyne, B doped and ortho BN co-doped $\gamma$-graphynes exhibit the rotoinversion (\(R_{\mathcal{P}} = C_{2z} + \mathcal{P}\)) and broken $\mathcal{PT}$ symmetry, resulting in non-zero $j_z^{ph}$. Furthermore, $\Gamma$-point phonon eigenvector analysis confirms that superposition of three $A$ modes gives rise to chiral phonon modes in both B doped and ortho BN co-doped $\gamma$-graphynes.

Figure \ref{fig:fig-5}a shows the phonon angular momentum values at K and K$'$ corresponding to the phonon bands for all doped graphyne configurations. Interestingly, at K and K$^{'}$ points, the chirality of the phonon modes is opposite to each other. This is because the wave vectors at these points are connected by $\mathcal{T}$. Although the number of phonon bands (or phonon frequency range) varies with number of atoms, a consistent trend is observed in $J{_z^{ph}}$ values for both doped 6-6-12 and $\gamma$-graphynes. This trend becomes particularly evident when analyzing the color bar intensities, especially at the maximum $J{_z^{ph}}$ (Figure \ref{fig:fig-5}a). In 6-6-12-graphyne, moving from B (electron-deficient) to BN (moderately electron-rich) to N (electron-rich) reveals a distinct increase in $|J{_z^{ph}}|$, following the order B $<$ BN $<$ N. The BN exhibits intermediate behavior, reflecting its balanced electron contribution. Interestingly, the phonon angular momentum values in B doped 6-6-12 and $\gamma$-graphynes remain nearly identical, owing to the nearly equivalent charge transfer from B to neighboring C atoms. Substitution from B to BN leads to a pronounced increase in $|J{_z^{ph}}|$ (Figures \ref{fig:fig-5}a,b), primarily due to electron enrichment in BN. However, BN co-doping in $\gamma$-graphyne induces higher phonon angular momentum compared to BN co-doped 6-6-12-graphyne, attributed to a larger charge transfer of $+$0.54 $e^-$ from BN to carbon environment (Figure \ref{fig:fig-5}c). On the other hand, BN co-doped 6-6-12-graphyne exhibits a lower charge transfer of $+$0.39 $e^-$ (Figure \ref{fig:fig-5}d), resulting in relatively lower phonon angular momentum values. This is because electron-rich dopants increase the local electron density, which could allow chiral phonon modes to interact more effectively with electronic environment~\cite{zhang2024effects}.

To gain deeper insight, we increase the BN co-dopant concentration and analyze its impact on phonon angular momentum. The optimized atomic geometries of 6-6-12 and $\gamma$-graphynes with higher BN co-doping are presented in Figures \ref{fig:fig-6}a,b, respectively. All non-equivalent bond lengths are clearly indicated in the inset of the structures. The corresponding phonon band structures are shown in Figure S5 in the SI, confirming their dynamical stability. Moreover, both configurations possess threefold rotational symmetry (C$_3$) while exhibiting broken $\mathcal{PT}$ symmetry. This leads to non-zero phonon angular momentum $J{_z^{ph}}$. The calculated maximum $|J{_z^{ph}}|$ at the time-reversal K point is depicted in Figure \ref{fig:fig-6}c. A decrease in $|J{_z^{ph}}|$ value with increasing BN concentration is observed for both graphyne variants, as indicated by the black dotted arrows.

To understand this trend, we compute the charge transfer between BN units and the surrounding carbon framework. Charge density difference plots reveal a transfer of 0.31 $e^-$ and 0.40 $e^-$ from BN to C atoms for 6-6-12 and $\gamma$-graphynes, respectively, under higher BN co-doping (Figures \ref{fig:fig-6}d,e). This enhanced charge transfer in $\gamma$-graphyne contributes to its relatively higher $|J{_z^{ph}}|$ compared to the 6-6-12-graphyne. However, when compared to the lower BN co-doping case (Figure \ref{fig:fig-5}a,b), the overall $|J{_z^{ph}}|$ values are reduced. This is consistent with reduced charge transfer at higher BN concentrations (Figure \ref{fig:fig-6}d,e), compared to the lower BN co-doping (Figure \ref{fig:fig-5}c,d) both for 6-6-12 and $\gamma$-graphynes. These findings further suggest that an electron-rich environment favors larger phonon angular momentum. In essence, electron-rich doping offers a promising pathway for tailoring chiral phonon behavior, which is crucial for designing advanced optoelectronic and phononic devices.

\section{Conclusion}
In summary, our work presents the first demonstration of the existence of chiral phonons in graphyne, achieved through atomic selective substitutional doping. The B, N doping and ortho BN co-doping break the $\mathcal{PT}$ symmetry in 6-6-12- and $\gamma$-graphynes. This, in turn, results in circularly polarized vibrational modes with quantized angular momentum at high-symmetry points (away from $\Gamma$ point). A correlation is observed between dopant electron affinity and chiral phonon mode of vibrations, indicating role of electron-phonon interaction in modulating phonon angular momentum. The proposed approach not only advances the fundamental understanding of chiral phonon in graphyne but also opens new avenues for designing phononic and optoelectronic devices.

\section*{ASSOCIATED CONTENT}

\section{Supporting Information}
Methods, Phonon band structure of $\beta$-graphyne, Various BN co-doped configurations and comparison of their relative energies, Phonon band structure of N doped $\gamma$-graphyne, Phonon band structures for higher BN co-doping, Correlation between symmetry and phonon angular momentum.

\section*{AUTHOR INFORMATION}

\section{Author Contributions}
A.K.S., P.A.S.A., and D.S.G. designed the project, supervised its execution, and provided overall guidance. S.M. performed all the calculations required for this work. A.C. helped with structure formation. S.M., A.C., D.S.G., P.A.S.A., and A.K.S. participated in the discussions and contributed with scientific input. S.M., A.C., D.S.G., P.A.S.A., and A.K.S. participated in manuscript writing. All authors have reviewed the manuscript and have given approval to the final version of the manuscript.

\section{Corresponding Authors}

\textbf{Abhishek K. Singh $-$} Materials Research Centre, Indian Institute of Science, Bangalore 560012, India;
Email-abhishek@iisc.ac.in

\textbf{Pedro AS Autreto $-$} Center for Natural and Human Sciences (CCNH), Federal University of ABC Rua Santa Ad´elia 166, Santo Andr´e 09210-170, Brazil;
Email-pedro.autreto@uafbc.edu.br

\textbf{Douglas S Galvao, $-$} Department of Applied Physics and Center for Computational Engineering and Sciences, State University of Campinas, Campinas, 13083-859, SP, Brazil
Email-galvao@ifi.unicamp.br

\section{Authors}
\textbf{Subhendu Mishra $-$} Materials Research Centre, Indian Institute of Science, Bangalore 560012, India

\textbf{Arpan Chakraborty $-$} Materials Research Centre, Indian Institute of Science, Bangalore 560012, India

\section*{Acknowledgement}
S.M., A.C., and A.K.S. thank the Materials Research Center (MRC), Solid State Structural and Chemistry Unit (SSCU), and Supercomputer Education and Research Center (SERC), Indian Institute of Science, for providing the computational facilities. S.M., A.C., and A.K.S. acknowledge support from the Institute of Eminence (IoE) scheme of the Ministry of Human Resource Development, Government of India. P.A.S.A. thanks the CNPq (Grant 308428/2022-6) and the National Institute of Science and Technology on Materials Informatics (grant no. 371610/2023-0). S.M., A.C., D.S.G., P.A.S.A., and A.K.S. also acknowledge the support from the "Scheme for Promotion of Academic and Research Collaboration (SPARC)" fellowship (Project code: SP 3571).


\begin{mcitethebibliography}{34}
\providecommand*\natexlab[1]{#1}
\providecommand*\mciteSetBstSublistMode[1]{}
\providecommand*\mciteSetBstMaxWidthForm[2]{}
\providecommand*\mciteBstWouldAddEndPuncttrue
  {\def\EndOfBibitem{\unskip.}}
\providecommand*\mciteBstWouldAddEndPunctfalse
  {\let\EndOfBibitem\relax}
\providecommand*\mciteSetBstMidEndSepPunct[3]{}
\providecommand*\mciteSetBstSublistLabelBeginEnd[3]{}
\providecommand*\EndOfBibitem{}
\mciteSetBstSublistMode{f}
\mciteSetBstMaxWidthForm{subitem}{(\alph{mcitesubitemcount})}
\mciteSetBstSublistLabelBeginEnd
  {\mcitemaxwidthsubitemform\space}
  {\relax}
  {\relax}

\bibitem[Zhang and Niu(2015)Zhang, and Niu]{zhang2015chiral}
Zhang,~L.; Niu,~Q. Chiral phonons at high-symmetry points in monolayer hexagonal lattices. \emph{Phys. Rev. Lett.} \textbf{2015}, \emph{115}, 115502\relax
\mciteBstWouldAddEndPuncttrue
\mciteSetBstMidEndSepPunct{\mcitedefaultmidpunct}
{\mcitedefaultendpunct}{\mcitedefaultseppunct}\relax
\EndOfBibitem
\bibitem[Zhu \latin{et~al.}(2018)Zhu, Yi, Li, Xiao, Zhang, Yang, Kaindl, Li, Wang, and Zhang]{zhu2018observation}
Zhu,~H.; Yi,~J.; Li,~M.-Y.; Xiao,~J.; Zhang,~L.; Yang,~C.-W.; Kaindl,~R.~A.; Li,~L.-J.; Wang,~Y.; Zhang,~X. Observation of chiral phonons. \emph{Science} \textbf{2018}, \emph{359}, 579--582\relax
\mciteBstWouldAddEndPuncttrue
\mciteSetBstMidEndSepPunct{\mcitedefaultmidpunct}
{\mcitedefaultendpunct}{\mcitedefaultseppunct}\relax
\EndOfBibitem
\bibitem[Ueda \latin{et~al.}(2023)Ueda, Garcia-Fernandez, Agrestini, Romao, van~den Brink, Spaldin, Zhou, and Staub]{ueda2023chiral}
Ueda,~H.; Garcia-Fernandez,~M.; Agrestini,~S.; Romao,~C.~P.; van~den Brink,~J.; Spaldin,~N.~A.; Zhou,~K.-J.; Staub,~U. Chiral phonons in quartz probed by X-rays. \emph{Nature} \textbf{2023}, \emph{618}, 946--950\relax
\mciteBstWouldAddEndPuncttrue
\mciteSetBstMidEndSepPunct{\mcitedefaultmidpunct}
{\mcitedefaultendpunct}{\mcitedefaultseppunct}\relax
\EndOfBibitem
\bibitem[Yin \latin{et~al.}(2021)Yin, Ulman, Liu, Granados~del {\'A}guila, Huang, Zhang, Serra, Sedmidubsky, Sofer, Quek, \latin{et~al.} others]{yin2021chiral}
Yin,~T.; Ulman,~K.~A.; Liu,~S.; Granados~del {\'A}guila,~A.; Huang,~Y.; Zhang,~L.; Serra,~M.; Sedmidubsky,~D.; Sofer,~Z.; Quek,~S.~Y.; others Chiral phonons and giant magneto-optical effect in CrBr3 2D magnet. \emph{Adv. Mater.} \textbf{2021}, \emph{33}, 2101618\relax
\mciteBstWouldAddEndPuncttrue
\mciteSetBstMidEndSepPunct{\mcitedefaultmidpunct}
{\mcitedefaultendpunct}{\mcitedefaultseppunct}\relax
\EndOfBibitem
\bibitem[Maity \latin{et~al.}(2022)Maity, Mostofi, and Lischner]{maity2022chiral}
Maity,~I.; Mostofi,~A.~A.; Lischner,~J. Chiral valley phonons and flat phonon bands in moir{\'e} materials. \emph{Phys. Rev. B} \textbf{2022}, \emph{105}, L041408\relax
\mciteBstWouldAddEndPuncttrue
\mciteSetBstMidEndSepPunct{\mcitedefaultmidpunct}
{\mcitedefaultendpunct}{\mcitedefaultseppunct}\relax
\EndOfBibitem
\bibitem[Chen \latin{et~al.}(2015)Chen, Zheng, Fuhrer, and Yan]{chen2015helicity}
Chen,~S.-Y.; Zheng,~C.; Fuhrer,~M.~S.; Yan,~J. Helicity-resolved Raman scattering of MoS$_2$, MoSe$_2$, WS$_2$, and WSe$_2$ atomic layers. \emph{Nano Lett.} \textbf{2015}, \emph{15}, 2526--2532\relax
\mciteBstWouldAddEndPuncttrue
\mciteSetBstMidEndSepPunct{\mcitedefaultmidpunct}
{\mcitedefaultendpunct}{\mcitedefaultseppunct}\relax
\EndOfBibitem
\bibitem[Kang \latin{et~al.}(2018)Kang, Jung, Shin, Sohn, Ryu, Kim, Hoesch, and Kim]{kang2018holstein}
Kang,~M.; Jung,~S.~W.; Shin,~W.~J.; Sohn,~Y.; Ryu,~S.~H.; Kim,~T.~K.; Hoesch,~M.; Kim,~K.~S. Holstein polaron in a valley-degenerate two-dimensional semiconductor. \emph{Nat. Mater.} \textbf{2018}, \emph{17}, 676--680\relax
\mciteBstWouldAddEndPuncttrue
\mciteSetBstMidEndSepPunct{\mcitedefaultmidpunct}
{\mcitedefaultendpunct}{\mcitedefaultseppunct}\relax
\EndOfBibitem
\bibitem[Li \latin{et~al.}(2019)Li, Wang, Jin, Lu, Lian, Meng, Blei, Gao, Taniguchi, Watanabe, \latin{et~al.} others]{li2019emerging}
Li,~Z.; Wang,~T.; Jin,~C.; Lu,~Z.; Lian,~Z.; Meng,~Y.; Blei,~M.; Gao,~S.; Taniguchi,~T.; Watanabe,~K.; others Emerging photoluminescence from the dark-exciton phonon replica in monolayer WSe2. \emph{Nat. Commun.} \textbf{2019}, \emph{10}, 2469\relax
\mciteBstWouldAddEndPuncttrue
\mciteSetBstMidEndSepPunct{\mcitedefaultmidpunct}
{\mcitedefaultendpunct}{\mcitedefaultseppunct}\relax
\EndOfBibitem
\bibitem[Li \latin{et~al.}(2019)Li, Wang, Jin, Lu, Lian, Meng, Blei, Gao, Taniguchi, Watanabe, \latin{et~al.} others]{li2019momentum}
Li,~Z.; Wang,~T.; Jin,~C.; Lu,~Z.; Lian,~Z.; Meng,~Y.; Blei,~M.; Gao,~M.; Taniguchi,~T.; Watanabe,~K.; others Momentum-dark intervalley exciton in monolayer tungsten diselenide brightened via chiral phonon. \emph{ACS nano} \textbf{2019}, \emph{13}, 14107--14113\relax
\mciteBstWouldAddEndPuncttrue
\mciteSetBstMidEndSepPunct{\mcitedefaultmidpunct}
{\mcitedefaultendpunct}{\mcitedefaultseppunct}\relax
\EndOfBibitem
\bibitem[Luo \latin{et~al.}(2023)Luo, Lin, Zhang, Chen, Blackert, Xu, Yakobson, and Zhu]{luo2023large}
Luo,~J.; Lin,~T.; Zhang,~J.; Chen,~X.; Blackert,~E.~R.; Xu,~R.; Yakobson,~B.~I.; Zhu,~H. Large effective magnetic fields from chiral phonons in rare-earth halides. \emph{Science} \textbf{2023}, \emph{382}, 698--702\relax
\mciteBstWouldAddEndPuncttrue
\mciteSetBstMidEndSepPunct{\mcitedefaultmidpunct}
{\mcitedefaultendpunct}{\mcitedefaultseppunct}\relax
\EndOfBibitem
\bibitem[Chaudhary \latin{et~al.}(2024)Chaudhary, Juraschek, Rodriguez-Vega, and Fiete]{chaudhary2024giant}
Chaudhary,~S.; Juraschek,~D.~M.; Rodriguez-Vega,~M.; Fiete,~G.~A. Giant effective magnetic moments of chiral phonons from orbit-lattice coupling. \emph{Phys. Rev. B} \textbf{2024}, \emph{110}, 094401\relax
\mciteBstWouldAddEndPuncttrue
\mciteSetBstMidEndSepPunct{\mcitedefaultmidpunct}
{\mcitedefaultendpunct}{\mcitedefaultseppunct}\relax
\EndOfBibitem
\bibitem[Carvalho \latin{et~al.}(2017)Carvalho, Wang, Mignuzzi, Roy, Terrones, Fantini, Crespi, Malard, and Pimenta]{carvalho2017intervalley}
Carvalho,~B.~R.; Wang,~Y.; Mignuzzi,~S.; Roy,~D.; Terrones,~M.; Fantini,~C.; Crespi,~V.~H.; Malard,~L.~M.; Pimenta,~M.~A. Intervalley scattering by acoustic phonons in two-dimensional MoS2 revealed by double-resonance Raman spectroscopy. \emph{Nat. commun.} \textbf{2017}, \emph{8}, 14670\relax
\mciteBstWouldAddEndPuncttrue
\mciteSetBstMidEndSepPunct{\mcitedefaultmidpunct}
{\mcitedefaultendpunct}{\mcitedefaultseppunct}\relax
\EndOfBibitem
\bibitem[Mishra \latin{et~al.}(2024)Mishra, Maity, and Singh]{mishra2024symmetry}
Mishra,~S.; Maity,~N.; Singh,~A.~K. Symmetry-assisted anomalous Hall conductivity in a CrS 2-CrBr 3 heterostructure. \emph{Phys. Rev. B} \textbf{2024}, \emph{110}, 125406\relax
\mciteBstWouldAddEndPuncttrue
\mciteSetBstMidEndSepPunct{\mcitedefaultmidpunct}
{\mcitedefaultendpunct}{\mcitedefaultseppunct}\relax
\EndOfBibitem
\bibitem[Barik \latin{et~al.}(2024)Barik, Mishra, Khazaei, Wang, Liang, Sun, Ranjbar, Tan, Wang, Yunoki, \latin{et~al.} others]{barik2024valley}
Barik,~R.~K.; Mishra,~S.; Khazaei,~M.; Wang,~S.; Liang,~Y.; Sun,~Y.; Ranjbar,~A.; Tan,~T.~L.; Wang,~J.; Yunoki,~S.; others Valley-polarized topological phases with in-plane magnetization. \emph{Nano Lett.} \textbf{2024}, \emph{24}, 13213--13218\relax
\mciteBstWouldAddEndPuncttrue
\mciteSetBstMidEndSepPunct{\mcitedefaultmidpunct}
{\mcitedefaultendpunct}{\mcitedefaultseppunct}\relax
\EndOfBibitem
\bibitem[Ohe \latin{et~al.}(2024)Ohe, Shishido, Kato, Utsumi, Matsuura, and Togawa]{ohe2024chirality}
Ohe,~K.; Shishido,~H.; Kato,~M.; Utsumi,~S.; Matsuura,~H.; Togawa,~Y. Chirality-induced selectivity of phonon angular momenta in chiral quartz crystals. \emph{Phys. Rev. Lett.} \textbf{2024}, \emph{132}, 056302\relax
\mciteBstWouldAddEndPuncttrue
\mciteSetBstMidEndSepPunct{\mcitedefaultmidpunct}
{\mcitedefaultendpunct}{\mcitedefaultseppunct}\relax
\EndOfBibitem
\bibitem[Uchida \latin{et~al.}(2008)Uchida, Takahashi, Harii, Ieda, Koshibae, Ando, Maekawa, and Saitoh]{uchida2008observation}
Uchida,~K.-I.; Takahashi,~S.; Harii,~K.; Ieda,~J.; Koshibae,~W.; Ando,~K.; Maekawa,~S.; Saitoh,~E. Observation of the spin Seebeck effect. \emph{Nature} \textbf{2008}, \emph{455}, 778--781\relax
\mciteBstWouldAddEndPuncttrue
\mciteSetBstMidEndSepPunct{\mcitedefaultmidpunct}
{\mcitedefaultendpunct}{\mcitedefaultseppunct}\relax
\EndOfBibitem
\bibitem[Kim \latin{et~al.}(2023)Kim, Vetter, Yan, Yang, Wang, Sun, Yang, Comstock, Li, Zhou, \latin{et~al.} others]{kim2023chiral}
Kim,~K.; Vetter,~E.; Yan,~L.; Yang,~C.; Wang,~Z.; Sun,~R.; Yang,~Y.; Comstock,~A.~H.; Li,~X.; Zhou,~J.; others Chiral-phonon-activated spin Seebeck effect. \emph{Nat. Mater.} \textbf{2023}, \emph{22}, 322--328\relax
\mciteBstWouldAddEndPuncttrue
\mciteSetBstMidEndSepPunct{\mcitedefaultmidpunct}
{\mcitedefaultendpunct}{\mcitedefaultseppunct}\relax
\EndOfBibitem
\bibitem[Pols \latin{et~al.}(2025)Pols, Brocks, Calero, and Tao]{pols2025chiral}
Pols,~M.; Brocks,~G.; Calero,~S.; Tao,~S. Chiral Phonons in 2D Halide Perovskites. \emph{Nano Lett.} \textbf{2025}, \relax
\mciteBstWouldAddEndPunctfalse
\mciteSetBstMidEndSepPunct{\mcitedefaultmidpunct}
{}{\mcitedefaultseppunct}\relax
\EndOfBibitem
\bibitem[Zhang \latin{et~al.}(2012)Zhang, Pei, and Wang]{zhang2012molecular}
Zhang,~Y.; Pei,~Q.; Wang,~C. A molecular dynamics investigation on thermal conductivity of graphynes. \emph{Computational Materials Science} \textbf{2012}, \emph{65}, 406--410\relax
\mciteBstWouldAddEndPuncttrue
\mciteSetBstMidEndSepPunct{\mcitedefaultmidpunct}
{\mcitedefaultendpunct}{\mcitedefaultseppunct}\relax
\EndOfBibitem
\bibitem[Baughman \latin{et~al.}(1987)Baughman, Eckhardt, and Kertesz]{Ray}
Baughman,~R.; Eckhardt,~H.; Kertesz,~M. Structure-property predictions for new planar forms of carbon: Layered phases containing sp 2 and sp atoms. \emph{The Journal of chemical physics} \textbf{1987}, \emph{87}, 6687--6699\relax
\mciteBstWouldAddEndPuncttrue
\mciteSetBstMidEndSepPunct{\mcitedefaultmidpunct}
{\mcitedefaultendpunct}{\mcitedefaultseppunct}\relax
\EndOfBibitem
\bibitem[Shao and Sun(2015)Shao, and Sun]{shao2015optical}
Shao,~Z.-G.; Sun,~Z.-L. Optical properties of $\alpha$-, $\beta$-, $\gamma$-, and 6, 6, 12-graphyne structures: first-principle calculations. \emph{Phys. E} \textbf{2015}, \emph{74}, 438--442\relax
\mciteBstWouldAddEndPuncttrue
\mciteSetBstMidEndSepPunct{\mcitedefaultmidpunct}
{\mcitedefaultendpunct}{\mcitedefaultseppunct}\relax
\EndOfBibitem
\bibitem[Peng \latin{et~al.}(2012)Peng, Ji, and De]{peng2012mechanical}
Peng,~Q.; Ji,~W.; De,~S. Mechanical properties of graphyne monolayers: a first-principles study. \emph{Phys. Chem. Chem. Phys.} \textbf{2012}, \emph{14}, 13385--13391\relax
\mciteBstWouldAddEndPuncttrue
\mciteSetBstMidEndSepPunct{\mcitedefaultmidpunct}
{\mcitedefaultendpunct}{\mcitedefaultseppunct}\relax
\EndOfBibitem
\bibitem[Majidi and Ayesh(2023)Majidi, and Ayesh]{majidi2023comparative}
Majidi,~R.; Ayesh,~A.~I. Comparative study of $\delta$-graphdiyne and $\delta$-graphyne: Insights into structural stability and electronic and optical properties. \emph{J. Phys. Chem. C.} \textbf{2023}, \emph{127}, 22234--22240\relax
\mciteBstWouldAddEndPuncttrue
\mciteSetBstMidEndSepPunct{\mcitedefaultmidpunct}
{\mcitedefaultendpunct}{\mcitedefaultseppunct}\relax
\EndOfBibitem
\bibitem[Desyatkin \latin{et~al.}(2022)Desyatkin, Martin, Aliev, Chapman, Fonseca, Galv{\~a}o, Miller, Stone, Wang, Zakhidov, \latin{et~al.} others]{Valentin1}
Desyatkin,~V.~G.; Martin,~W.~B.; Aliev,~A.~E.; Chapman,~N.~E.; Fonseca,~A.~F.; Galv{\~a}o,~D.~S.; Miller,~E.~R.; Stone,~K.~H.; Wang,~Z.; Zakhidov,~D.; others Scalable synthesis and characterization of multilayer $\gamma$-graphyne, new carbon crystals with a small direct band gap. \emph{J. Am. Chem. Soc.} \textbf{2022}, \emph{144}, 17999--18008\relax
\mciteBstWouldAddEndPuncttrue
\mciteSetBstMidEndSepPunct{\mcitedefaultmidpunct}
{\mcitedefaultendpunct}{\mcitedefaultseppunct}\relax
\EndOfBibitem
\bibitem[Aliev \latin{et~al.}(2025)Aliev, Guo, Fonseca, Razal, Wang, Galv{\~a}o, Bolding, Chapman-Wilson, Desyatkin, Leisen, \latin{et~al.} others]{Valentin2}
Aliev,~A.~E.; Guo,~Y.; Fonseca,~A.~F.; Razal,~J.~M.; Wang,~Z.; Galv{\~a}o,~D.~S.; Bolding,~C.~M.; Chapman-Wilson,~N.~E.; Desyatkin,~V.~G.; Leisen,~J.~E.; others A planar-sheet nongraphitic zero-bandgap sp2 carbon phase made by the low-temperature reaction of $\gamma$-graphyne. \emph{Proc. Natl. Acad. Sci.} \textbf{2025}, \emph{122}, e2413194122\relax
\mciteBstWouldAddEndPuncttrue
\mciteSetBstMidEndSepPunct{\mcitedefaultmidpunct}
{\mcitedefaultendpunct}{\mcitedefaultseppunct}\relax
\EndOfBibitem
\bibitem[Malko \latin{et~al.}(2012)Malko, Neiss, Vines, and G{\"o}rling]{malko2012competition}
Malko,~D.; Neiss,~C.; Vines,~F.; G{\"o}rling,~A. Competition for graphene: graphynes with direction-dependent dirac cones. \emph{Phys. Rev. Lett.} \textbf{2012}, \emph{108}, 086804\relax
\mciteBstWouldAddEndPuncttrue
\mciteSetBstMidEndSepPunct{\mcitedefaultmidpunct}
{\mcitedefaultendpunct}{\mcitedefaultseppunct}\relax
\EndOfBibitem
\bibitem[Ali \latin{et~al.}(2025)Ali, Starczewska, Das, and Jesionek]{ali2025exploration}
Ali,~M.~D.; Starczewska,~A.; Das,~T.~K.; Jesionek,~M. Exploration of Sp-Sp$^2$ Carbon Networks: Advances in Graphyne Research and Its Role in Next-Generation Technologies. \emph{Int. J. Mol. Sci.} \textbf{2025}, \emph{26}, 5140\relax
\mciteBstWouldAddEndPuncttrue
\mciteSetBstMidEndSepPunct{\mcitedefaultmidpunct}
{\mcitedefaultendpunct}{\mcitedefaultseppunct}\relax
\EndOfBibitem
\bibitem[Perkg{\"o}z and Sevik(2014)Perkg{\"o}z, and Sevik]{perkgoz2014vibrational}
Perkg{\"o}z,~N.~K.; Sevik,~C. Vibrational and thermodynamic properties of $\alpha$-, $\beta$-, $\gamma$-, and 6, 6, 12-graphyne structures. \emph{Nanotechnology} \textbf{2014}, \emph{25}, 185701\relax
\mciteBstWouldAddEndPuncttrue
\mciteSetBstMidEndSepPunct{\mcitedefaultmidpunct}
{\mcitedefaultendpunct}{\mcitedefaultseppunct}\relax
\EndOfBibitem
\bibitem[Zhang and Niu(2014)Zhang, and Niu]{zhang2014angular}
Zhang,~L.; Niu,~Q. Angular momentum of phonons and the Einstein--de Haas effect. \emph{Phys. Rev. Lett.} \textbf{2014}, \emph{112}, 085503\relax
\mciteBstWouldAddEndPuncttrue
\mciteSetBstMidEndSepPunct{\mcitedefaultmidpunct}
{\mcitedefaultendpunct}{\mcitedefaultseppunct}\relax
\EndOfBibitem
\bibitem[Zhang \latin{et~al.}(2025)Zhang, Peshcherenko, Yang, Ward, Raghuvanshi, Lindsay, Felser, Zhang, Yan, and Miao]{zhang2025measurement}
Zhang,~H.; Peshcherenko,~N.; Yang,~F.; Ward,~T.; Raghuvanshi,~P.; Lindsay,~L.; Felser,~C.; Zhang,~Y.; Yan,~J.-Q.; Miao,~H. Measurement of phonon angular momentum. \emph{Nature Physics} \textbf{2025}, 1--5\relax
\mciteBstWouldAddEndPuncttrue
\mciteSetBstMidEndSepPunct{\mcitedefaultmidpunct}
{\mcitedefaultendpunct}{\mcitedefaultseppunct}\relax
\EndOfBibitem
\bibitem[Yang \latin{et~al.}(2025)Yang, Xiao, Mao, Li, Wang, Deng, Tang, Song, Li, Yuan, \latin{et~al.} others]{yang2025catalogue}
Yang,~Y.; Xiao,~Z.; Mao,~Y.; Li,~Z.; Wang,~Z.; Deng,~T.; Tang,~Y.; Song,~Z.-D.; Li,~Y.; Yuan,~H.; others Catalogue of chiral phonon materials. \emph{arXiv preprint arXiv:2506.13721} \textbf{2025}, \relax
\mciteBstWouldAddEndPunctfalse
\mciteSetBstMidEndSepPunct{\mcitedefaultmidpunct}
{}{\mcitedefaultseppunct}\relax
\EndOfBibitem
\bibitem[Zhang \latin{et~al.}(2025)Zhang, Huang, Du, Ying, Du, and Zhang]{zhang2025chirality}
Zhang,~S.; Huang,~Z.; Du,~M.; Ying,~T.; Du,~L.; Zhang,~T. The chirality of phonons: Definitions, symmetry constraints, and experimental observation. \emph{arXiv preprint arXiv:2503.22794} \textbf{2025}, \relax
\mciteBstWouldAddEndPunctfalse
\mciteSetBstMidEndSepPunct{\mcitedefaultmidpunct}
{}{\mcitedefaultseppunct}\relax
\EndOfBibitem
\bibitem[Zhang \latin{et~al.}(2024)Zhang, Wang, and Lian]{zhang2024effects}
Zhang,~J.; Wang,~F.; Lian,~C.-S. Effects of charge doping and interfacial interaction on the charge density wave order in monolayer 1 T-TiTe 2 and 1 T-ZrTe 2. \emph{Phys. Rev. B} \textbf{2024}, \emph{110}, 245416\relax
\mciteBstWouldAddEndPuncttrue
\mciteSetBstMidEndSepPunct{\mcitedefaultmidpunct}
{\mcitedefaultendpunct}{\mcitedefaultseppunct}\relax
\EndOfBibitem
\end{mcitethebibliography}
\providecommand{\latin}[1]{#1}
\makeatletter
\providecommand{\doi}
  {\begingroup\let\do\@makeother\dospecials
  \catcode`\{=1 \catcode`\}=2 \doi@aux}
\providecommand{\doi@aux}[1]{\endgroup\texttt{#1}}
\makeatother
\providecommand*\mcitethebibliography{\thebibliography}
\csname @ifundefined\endcsname{endmcitethebibliography}  {\let\endmcitethebibliography\endthebibliography}{}

\end{document}